# The relativistic Doppler effect: when a zero frequency shift or a red shift exists for sources approaching the observer

**Changbiao Wang***
ShangGang Group, 70 Huntington Road, Apartment 11, New Haven, CT 06512, USA

It is shown without making use of Lorentz transformation that there exists a phenomenon of relativistic zero-frequency shift in Doppler effect for a plane wave in free space, observed in two inertial frames of relative motion, and the zero shift takes place at a maximum aberration of light. When it is applied to analysis of a moving point light source, two unconventional physical implications result: (1) a light source, when it is approaching (moving closer to) the observer, may cause a red shift; (2) a zero-frequency-shift observation does not necessarily mean that the light source is not moving closer, and in contrast, the light source may be moving closer to the observer at a high speed. This fundamental result of special relativity may provide an alternative way to experimentally examine the principle of relativity, and might have a significant application in astrophysics.

Principle of relativity and constancy of the light speed in free space are the two basic postulates of the special theory of relativity [1,2]. A uniform plane electromagnetic wave, which is a fundamental solution to Maxwell equations, propagates at the light speed in all directions [3]. No observers can identify whether this plane wave is in motion or not, although its frequency, propagation direction, and field strength can be measured. Consequently, when directly applying the relativity principle to Maxwell equations, the light speed must be the same in all inertial frames of reference, in other words, the covariance of Maxwell equations requires the constancy of light speed. Therefore, Einstein's first and second postulates are compatible [4-7].

It is a well-known statement in the physics community that a moving light source causes a blue shift for approaching and a red shift for receding [8]. The relativistic transverse Doppler effect is an important result of the special theory of relativity [1] and has been demonstrated experimentally [9,10]; this effect occurs at a critical point for approaching and receding, but it is a red shift. Furthermore, Hovsepyan has shown from the geometric standpoint that, the red shift may exist even when the light source really approaches the observer [11]. All these results clearly show contradiction with the above well-known statement. Behind the contradiction, there must be some physics that has not been exposed.

In this paper, an intuitive but rigorous proof is given of the existence of relativistic zero-frequency shift in Doppler effect for a plane wave in free space, observed in two inertial frames of relative motion, and the zero shift taking place at a maximum aberration of light. The red shift for approaching is a significant implication of the strict theoretic result, although it is against the conventional viewpoint. A conceptual experiment for verifying this phenomenon is suggested.

We begin with an instructive derivation of relativistic Doppler and aberration formulas based on an infinite uniform electromagnetic plane wave in free space without making use of Lorentz transformation [1]. First let us examine the properties of the plane wave. According to the relativity principle, the plane wave in any inertial frame has a phase factor $\exp i(\omega t - \mathbf{k}\cdot\mathbf{r})$, where $t$ is the time, $\omega$ is the frequency, $\mathbf{r}$ is the position vector in space, and $|\mathbf{k}| = \omega/c$ is the wave number, with $c$ the vacuum light speed. According to the phase invariance [1,12], the phase $\psi = \omega t - \mathbf{k}\cdot\mathbf{r}$ takes the same value in all inertial frames for a given *time-space point*. If $\psi_1$ is the phase at the first time-space point where the wave reaches its crest and $\psi_2$ is the one at the second such point, with $|\psi_2 - \psi_1| = 2\pi$, then the two crest-time-space points are said to be "successive", and $|\omega\Delta t - \mathbf{k}\cdot\Delta\mathbf{r}| = 2\pi$ holds in *all inertial frames*, where $\Delta t$ and $\Delta\mathbf{r}$ are, respectively, the differences between the two time-space points.

* Email: changbiao_wang@yahoo.com

Observed at the *same time*, the set of all the space points satisfying $\omega t - \mathbf{k}\cdot\mathbf{r} = \psi =$ constant is defined as the wavefront, which is an equiphase plane with the wave vector $\mathbf{k}$ as its normal, and moves at $c$ along the $\mathbf{k}$-direction. Obviously, observed at the same time, two successive crest-wavefronts are "adjacent" geometrically.

Now let us give the definitions of wave period and wavelength in terms of the expression $|\omega\Delta t - \mathbf{k}\cdot\Delta\mathbf{r}| = 2\pi$ given above. In a given inertial frame, observed at the *same point* $(\Delta\mathbf{r} = 0)$, the time difference $\Delta t$ between the occurrences of two successive crest-wavefronts is defined to be the wave period $T = \Delta t = 2\pi/\omega$; observed at the *same time* ($\Delta t = 0$), the space distance between two adjacent crest-wavefronts, given by $|\Delta\mathbf{r}|$ with $\Delta\mathbf{r} \,//\, \mathbf{k}$, is defined to be the wavelength $\lambda = |\Delta\mathbf{r}| = 2\pi/|\mathbf{k}| = cT = 2\pi c/\omega$.

From above we see that the definitions of wave period and wavelength are closely associated with time and space. Suppose that one observer is fixed at the origin $O$ of the $XOY$ frame, and the other is fixed at the origin $O'$ of the $X'O'Y'$ frame, which moves relatively to $XOY$ at a velocity of $v = \beta c$ along the $x$-direction. All corresponding axes of the two frames have the same directions. Observed in the $XOY$ frame at the instant $t = t_1$, there are two successive (adjacent) crest-wavefronts, with the $O'$-observer overlapping $O_1'$ on the first wavefront. Observed at the instant $t = t_2$ $(> t_1)$, the two crest-wavefronts are located in a new place, with the $O'$-observer overlapping $O_2'$ on the second wavefront; as shown in Fig. 1. The distance between the two crest-wavefronts, measured by the $O$-observer, is one wavelength ($\lambda$). From Fig. 1, we have

$$t_2 = t_1 + \lambda/c + O_1'O_2'\cos\phi/c\,. \tag{1}$$

Inserting $\lambda = cT$ and $O_1'O_2' = v(t_2 - t_1)$ into above, we have

$$(t_2 - t_1)(1 - \mathbf{n}\cdot\boldsymbol{\beta}) = T\,, \tag{2}$$

where $\mathbf{n}\cdot\boldsymbol{\beta} = \beta\cos\phi$, with $\mathbf{n} = \mathbf{k}/|\mathbf{k}|$ the unit wave vector, and $\beta = |\boldsymbol{\beta}| = |\mathbf{v}/c|$.

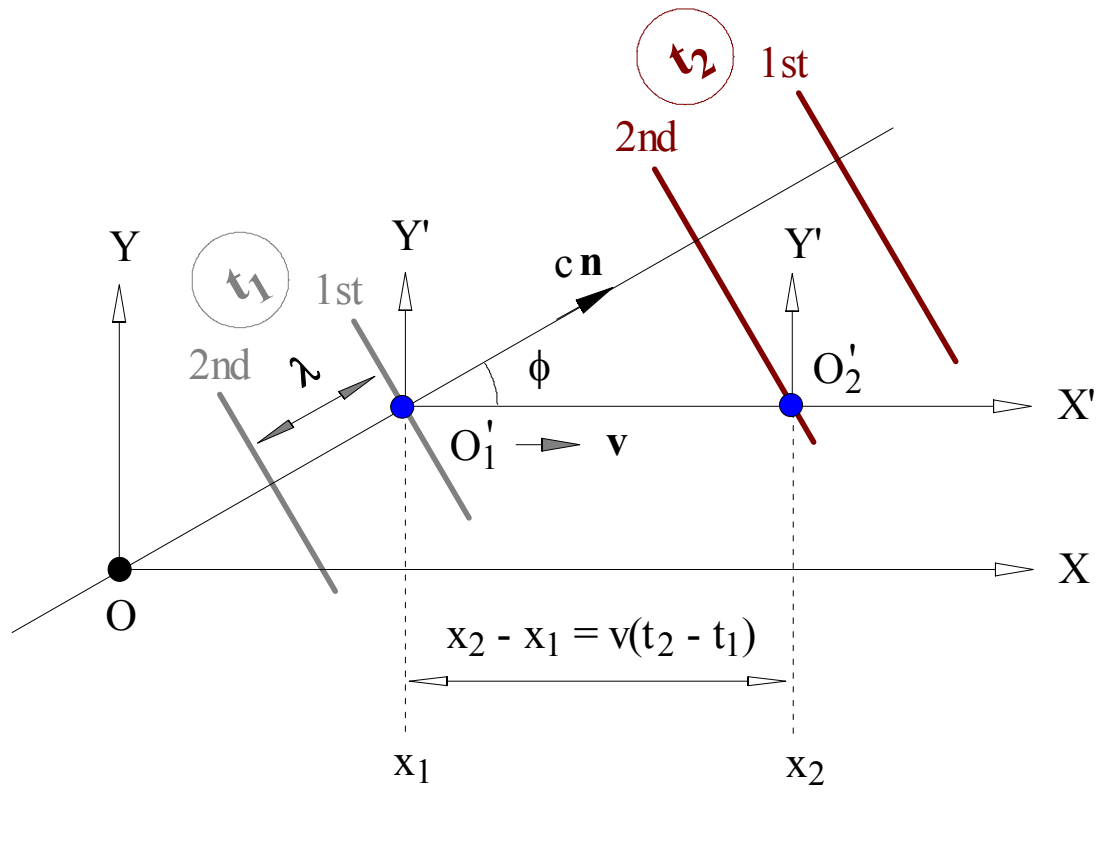

Fig. 1. Two adjacent crest-wavefronts of a plane wave in free space, which are observed in the $XOY$ frame respectively at $t = t_1$ and $t_2$. $O'$-observer moves relatively to $O$-observer at a velocity of $v$ along the $x$-direction. At $t_1$, the $O'$-observer overlaps $O_1'$ on the 1st wavefront; at $t_2$, the $O'$-observer overlaps $O_2'$ on the 2nd wavefront.

Observed in the $X'O'Y'$ frame, the two successive crest-wavefronts, which are adjacent in the $XOY$ frame, both sweep over the $O'$-observer at the same place ($\Delta\mathbf{r}' = 0$). According to the phase invariance [1,12], we have $|\omega'\Delta t' - \mathbf{k}'\cdot\Delta\mathbf{r}'| = |\omega'\Delta t'| = 2\pi$, or $\omega'(t_2' - t_1') = 2\pi$. Thus we have the wave period in the

$X'O'Y'$ frame, given by $T' = t_2' - t_1' = 2\pi/\omega'$ in terms of the definition given previously. Due to the time dilation, we have $t_2 - t_1 = \gamma(t_2' - t_1') = \gamma(2\pi/\omega')$, where $\gamma = (1-\beta^2)^{-1/2}$ is the time dilation factor. Inserting $t_2 - t_1 = \gamma(2\pi/\omega')$ and $T = 2\pi/\omega$ into Eq. (2), we have the Doppler formula [1], given by

$$\omega' = \omega\gamma(1 - \mathbf{n}\cdot\boldsymbol{\beta}) . \tag{3}$$

It should be noted that Eq. (2) is pure classical, while Eq. (3) is relativistic with the time dilation effect taken into account in the derivation. Since the reciprocity principle holds in special relativity [13], we may assume that $XOY$ frame moves relatively to $X'O'Y'$ frame at a velocity of $\mathbf{v}' = -\mathbf{v}$ along the minus $x'$-direction. A similar derivation yields [1]

$$\omega = \omega'\gamma'(1 - \mathbf{n}'\cdot\boldsymbol{\beta}') , \tag{4}$$

where $\mathbf{n}' = \mathbf{k}'/|\mathbf{k}'|$ with $|\mathbf{k}'| = \omega'/c$, $\boldsymbol{\beta}' = -\boldsymbol{\beta}$ with $\beta' = \beta$, and $\gamma' = \gamma$.

Inserting Eq. (3) into Eq. (4), we obtain the formula for measuring aberration of light [1], given by

$$\boldsymbol{\beta}'\cdot\mathbf{n}' = \frac{\boldsymbol{\beta}^2 - \boldsymbol{\beta}\cdot\mathbf{n}}{1 - \boldsymbol{\beta}\cdot\mathbf{n}} , \tag{5}$$

or

$$\cos\phi' = \frac{\beta - \cos\phi}{1 - \beta\cos\phi} , \tag{6}$$

where $\phi$ is the angle between $\boldsymbol{\beta}$ and $\mathbf{n}$, and $\phi'$ is the one between $\boldsymbol{\beta}'$ and $\mathbf{n}'$; both limited in the range of $0 \le \phi, \phi' \le \pi$. Because of the aberration of light, $\phi + \phi' \le \pi$ must hold and the equal sign is valid only for $\beta = 0$, $\phi = 0$ or $\pi$. Since no observers can identify whether the plane wave in free space is in motion or not, a light aberration is relative and it is convenient to use $\phi + \phi'$ to measure the aberration. When $\phi + \phi' = \pi$, there is no aberration; when $\phi + \phi' < \pi$, there is an aberration.

It should be emphasized that Eqs. (3)-(6) are independent of the choice of inertial frames, and the primed and unprimed quantities, as illustrated in Fig. 2, are exchangeable. From Eqs. (3) and (4), we have

$$\omega' = \omega\sqrt{\frac{1 - \beta\cos\phi}{1 - \beta\cos\phi'}} \begin{cases} > \omega, & \text{if } \phi' < \phi \\ = \omega, & \text{if } \phi' = \phi \\ < \omega, & \text{if } \phi' > \phi \end{cases} . \tag{7}$$

From the above Eq. (7) we find $\omega' = \omega$ when the two position angles are equal ($\phi' = \phi$), which means that there is no frequency shift in such case although the light aberration must exist ($\phi + \phi' \ne \pi$ for $\phi' = \phi$ and $\beta \ne 0$). Setting $\phi = \phi'$ in Eq. (6), we obtain the condition for the Doppler zero-frequency shift, given by

$$\phi_{zfs} = \cos^{-1}\sqrt{\frac{\gamma - 1}{\gamma + 1}} , \qquad (0 \le \beta < 1) , \tag{8}$$

or

$$\beta = \frac{2\cos\phi_{zfs}}{1 + \cos^2\phi_{zfs}} . \tag{9}$$

In other words, if $XOY$-observer has a position angle $\phi = \phi_{zfs}$ for a plane wave in free space, then the $X'O'Y'$-observer, who moves relatively to the $XOY$-observer at the normalized velocity given by Eq. (9), has an equal position angle and the same observed frequency (confer Fig. 2).

Note: $\phi_{zsf} < 0.5\pi$ holds for $\beta \ne 0$, $\phi_{zfs} \approx 0.5(\pi - \beta)$ for $\gamma \approx 1$ ($\beta << 1$), and $\phi_{zfs} \approx (2/\gamma)^{1/2}$ for $\gamma >> 1$ ($\beta \approx 1$). As a numerical example, the light aberration and Doppler effect are shown in Fig. 3 for $\gamma = 2$ ($\beta = 0.8660$), with the zero shift taking place at $\phi = \phi_{zfs} = 0.304\pi = 54.7^{\circ}$, where the aberration reaches maximum.

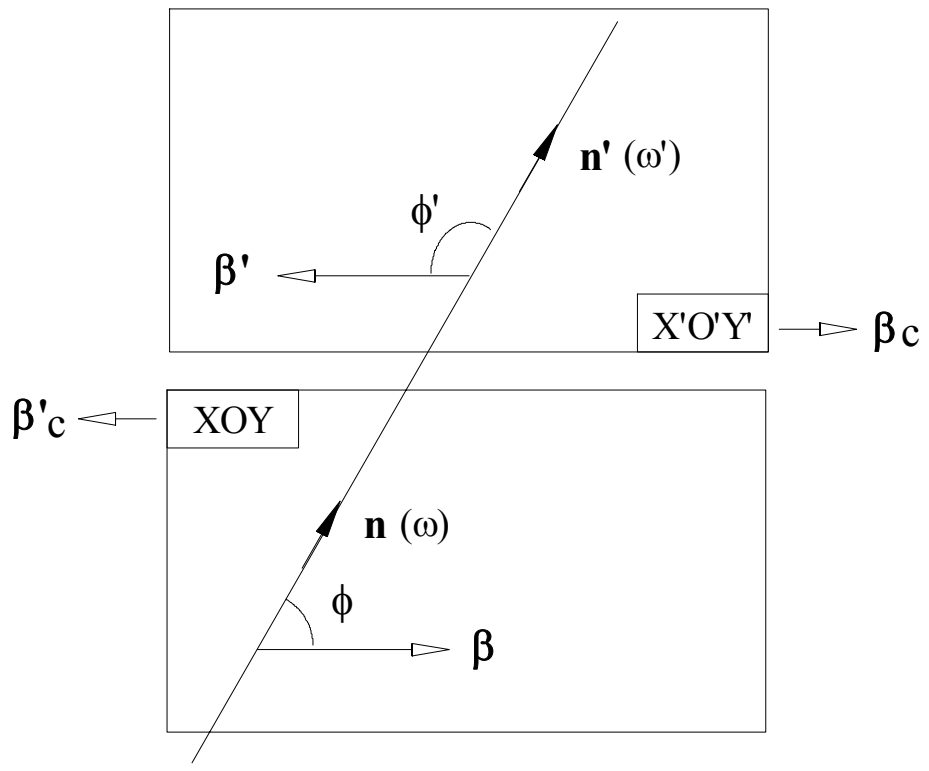

Fig. 2. A plane wave in free space observed in inertial frames $XOY$ and $X'O'Y'$ which are in relative motion. $\boldsymbol{\beta}c$ is the velocity of $X'O'Y'$ relative to $XOY$, and $\boldsymbol{\beta}'c$ is the velocity of $XOY$ relative to $X'O'Y'$. $\mathbf{n}$ and $\mathbf{n}'$ are the unit wave vectors, and $\omega$ and $\omega'$ are the frequencies, respectively measured in the two frames. Transverse Doppler effect: (*a*) $\omega' = \gamma\omega$ and $\cos\phi' = \beta$ for $\phi = \pi/2$; (*b*) $\omega = \gamma\omega'$ and $\cos\phi = \beta' = \beta$ for $\phi' = \pi/2$. Doppler zero shift: $\omega' = \omega$ at $\phi' = \phi = \phi_{zfs}$.

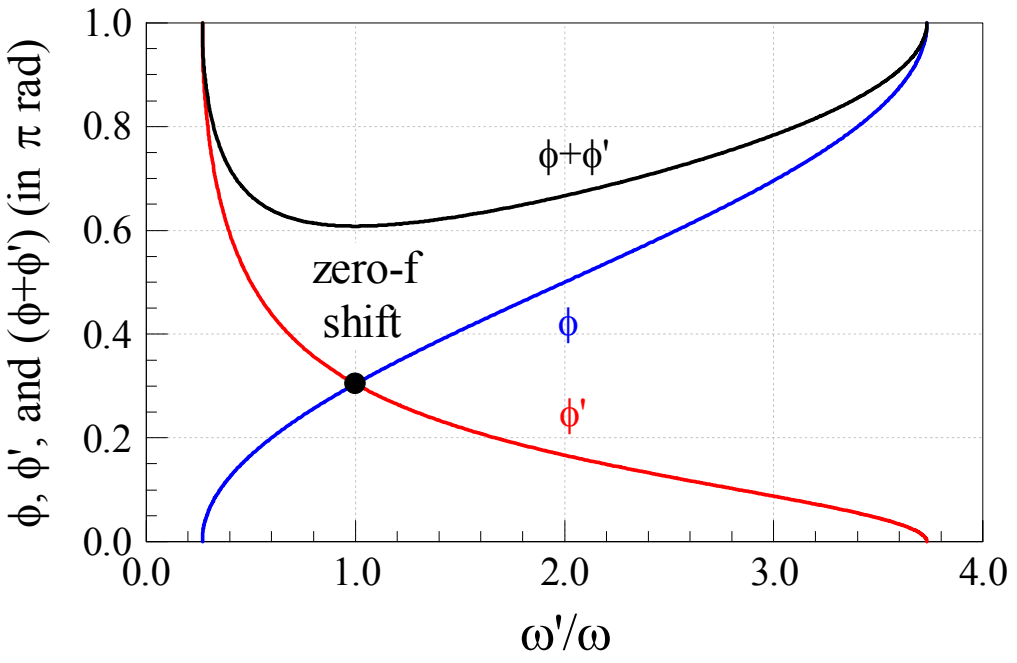

Fig. 3. Light aberration and Doppler frequency shift for a plane wave in free space observed in two inertial frames, which are in relative motion with a velocity of $\beta c$. $\phi + \phi' = \pi$ corresponds to no aberration. The zero-shift angle $\phi = \phi' = \phi_{zfs}$ is marked with a solid dot, where $\phi + \phi'$ reaches minimum, corresponding to a maximum aberration. $\omega'/\omega < 1$ for $\phi < \phi_{zfs}$, $\omega'/\omega = 1$ for $\phi = \phi_{zfs}$, and $\omega'/\omega > 1$ for $\phi > \phi_{zfs}$.

Due to the relativity of motion, Doppler effect is symmetric; for example, from $\omega'/\omega < 1$ for $\phi < \phi_{zfs}$, by exchanging the primed and unprimed quantities we have $\omega/\omega' < 1$ for $\phi' < \phi_{zfs}$, as seen in Fig. 3 [14]. When applied to analysis of a light source that moves relatively to an observer, any one of $\omega$ and $\omega'$ can be the frequency of the light source. For the convenience of specific description below, $\omega'$ is taken to be the frequency of a moving light source and $\omega$ is taken to be the observer's frequency.

The relativistic Doppler zero shift has an unusual physical implication: a light source, when it is moving closer to the observer, may produce not only a blue shift but also a red shift; in other words, a red shift is not necessarily to give an explanation that the light source is receding away. As illustrated in Fig. 4, for $0 \le \phi < \phi_{zfs}$ the observer's frequency changes from $\omega = \omega'[(1+\beta)/(1-\beta)]^{1/2}$ to $\omega'$, which is blue-shifted. If $\gamma$ is very large, then the blue shift is limited within a relatively small angle range, namely $0 \le \phi < (2/\gamma)^{1/2}$ with $2\gamma\omega' \ge \omega > \omega'$. A red shift $\omega < \omega'$ takes place when $\phi > \phi_{zfs}$. In the region of $\phi_{zfs} < \phi < \pi/2$, the light source is approaching the observer, with $\omega' > \omega > \omega'/\gamma$, so-called "red shift for approaching", which is a relativistic effect.

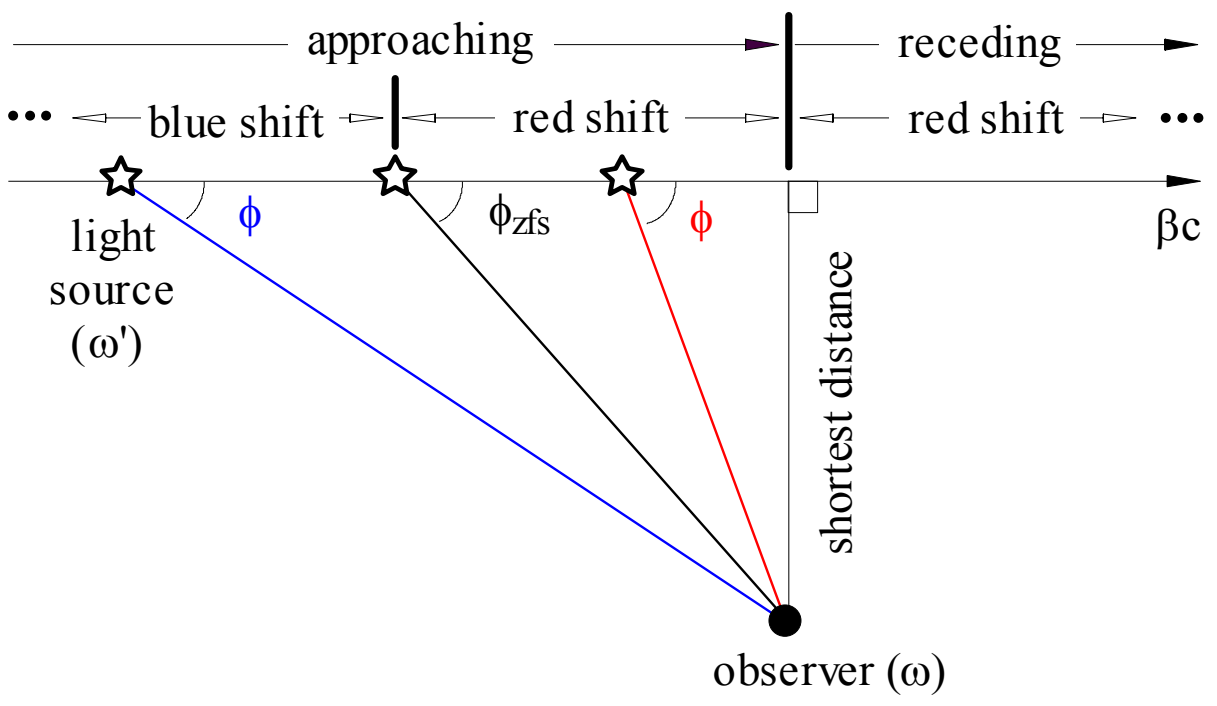

Fig. 4. Illustration of the existence of red shift when a light source is moving at $\beta c$ closer to the observer. $\omega > \omega'$ for $\phi < \phi_{zfs}$ (blue shift), and $\omega < \omega'$ for $\phi > \phi_{zfs}$ (red shift). In the region of $\phi_{zfs} < \phi < \pi/2$ where the light source is approaching the observer, the frequency observed is red-shifted with $\omega' > \omega > \omega'/\gamma$ .

Exact longitudinal Doppler effect is not easy to get in practice while any transverse effect $(0 < \phi < \pi/2)$ may allow the existence of "red shift for approaching". A red shift, measured without knowing the exact direction in which a light source moves, may be explained to be the light source's moving not only "away from" but also "closer to" the observer. Such a specific example is shown in Fig. 5.

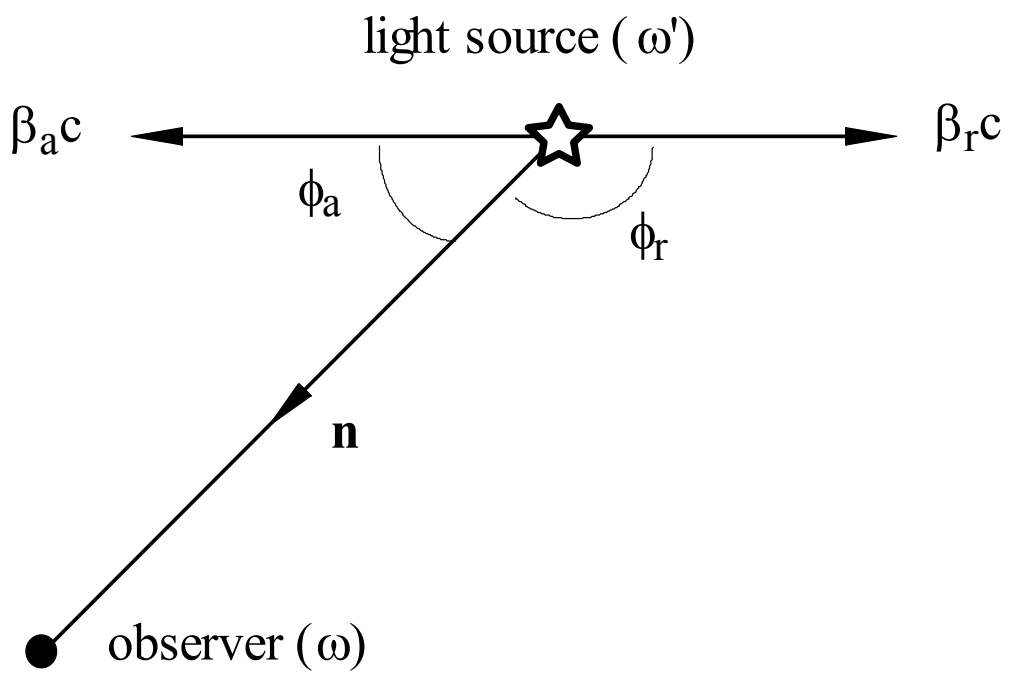

Fig. 5. Examples of red shift for receding and approaching. Suppose that a red shift is known with $\omega'/\omega = 1.1$ but the observer does not know the exact direction along which the light source moves. Thus the observer can take the light source to be receding away at $\beta_r = 0.1285$ with $\phi_r = 3\pi/4$ ; however, also can take it to be approaching at $\beta_a = 0.9555$ with $\phi_a = \pi/4$ .

Some important differences between the relativistic and classic Doppler effects are outlined as follows.

(1) In the relativistic Doppler effect, the zero-shift angle changes in the range of $\pi/2 > \phi_{zfs} > 0$ , depending on the relative velocity between the moving light source and the observer. Since the zero-shift angle $\phi_{zfs} < \pi/2$ holds for any $\beta \neq 0$ , a zero-shift observation does not necessarily mean that the light source is not moving closer; in contrast, the light source may be moving closer to the observer at a large velocity. In the classic Doppler effect, however, the zero-shift angle is always equal to $\pi/2$ , independently of the relative velocity [3], and a zero-shift observation exactly corresponds to the light source's not moving closer to the observer.

(2) For the relativistic Doppler effect, a given quantity of red shift does not correspond to a unique radial velocity, $c\beta\cos\phi$ , as seen in Eq. (3), and this red shift may correspond to an infinite number of approaching or receding velocities. For the classic Doppler effect, however, the frequency shift and the radial velocity $c\beta\cos\phi$ are one-to-one corresponding, and the red shift takes place only when the light source is receding away from the observer.

One may use a moving electron bunch in a free-electron laser [15] as a moving light source to experimentally verify this zero-shift angle. For a moving light source made by a 24-MeV electron bunch [16], the angle $\phi_{zfs}$ is only about 11.7 degrees, and a suggested conceptual experiment arrangement is shown in Fig. 6. The bunch's proper radiation frequency cannot be directly measured, but it can be obtained from theory and then compared with the one measured at the zero-shift angle. The main difficulty for such an experiment is probably the requirement of high sensitivity detector for detecting radiation power because the most power is distributed within a small angle of $\phi \sim 1/\gamma$ [3], which is smaller than the zero-shift angle $\phi_{zfs} \approx (2/\gamma)^{1/2}$.

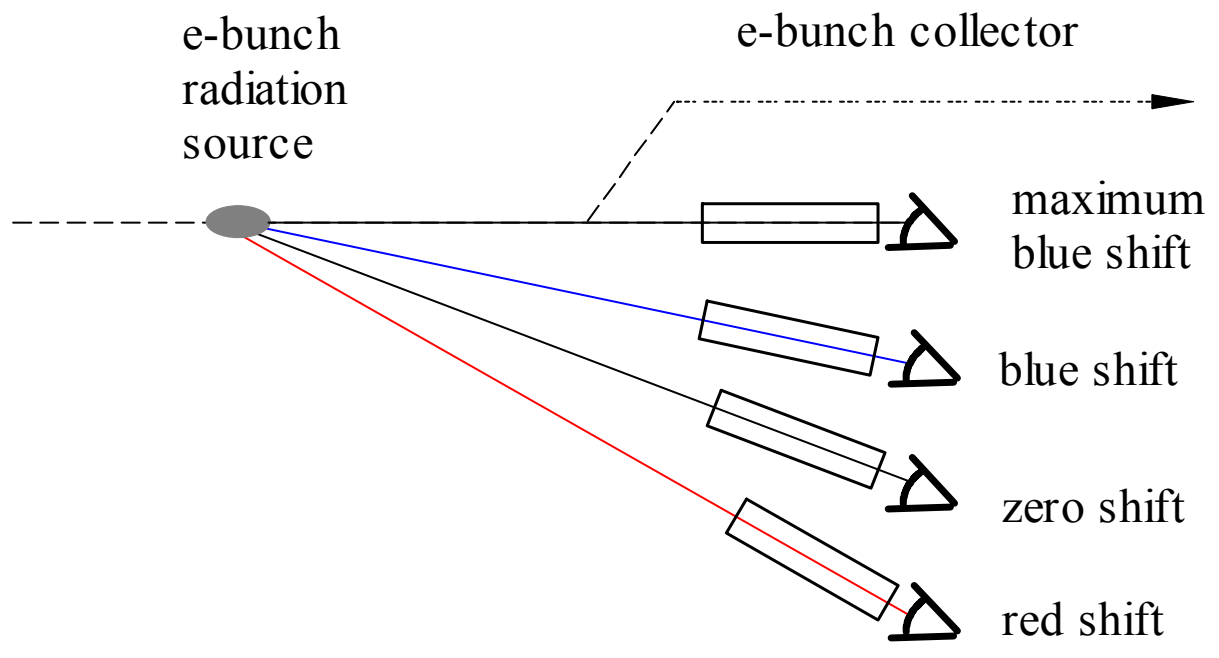

Fig. 6. A suggested conceptual experiment arrangement for verifying the zero-shift angle and a red shift for approaching. A space-periodically-modulated electron bunch [15,16] is used as a moving light source, and the radiation frequency is observed at different angles. By comparing the frequency measured at the zero-shift angle with the bunch modulation frequency, which differs by a time-dilation factor theoretically, the time dilation effect also can be checked.

It should be pointed out that Eq. (8) or Eq. (9) is a strict result of the special relativity for a uniform plane wave in free space, which is observed in two inertial frames in relative motion. Under this zero-shift condition, observed in the two frames respectively, the electric or magnetic field amplitudes of the plane wave are equal [1], with a maximum aberration of light. However any existing light source or a theoretical point light source may only generate a "local" plane wave, and the application of Eq. (8) or Eq. (9) is an approximation of the strict theoretic result.

Doppler effect is often used for studying motions of celestial bodies, and the Doppler zero shift might have an important application in astrophysics. For example, it is well recognized that light from most galaxies is Doppler-red-shifted, which is usually explained in university physics textbooks to be these galaxies' moving *away from* us [8]. Since there may be a relativistic red shift for a light source to move *closer to* us, as shown in the paper, the above explanation probably should be revised. To show this, an illustrative example is given below.

Suppose that a distant galaxy, which has a shape of oblate ellipsoid with a dimension of $2\times10^5$ light years and is $10^9$ light years away, moves towards Earth at a nearly light speed ($\gamma = 5\times10^8$), as shown in Fig. 7. All the electromagnetic radiations observed on the Earth are distributed within a small angle of $2\phi_b \approx 2\times10^{-4}$ rad, almost parallel. From Eq. (3), we have

$$\lambda/\lambda' \approx (1+\gamma^2\phi^2)/(2\gamma) \qquad \text{for } \gamma >> 1 \text{ and } \phi \approx 0,$$

leading to $1/(2\gamma) < \lambda/\lambda' < \gamma\phi_b^2/2$, with $10^{-9} < \lambda/\lambda' < 1$ in the blue-shift regime $0 \le \phi < \phi_{zfs}$, and $1 < \lambda/\lambda' < 2.5$ in the red-shift regime $\phi_{zfs} < \phi \le \phi_b$, where $\phi_{zfs} \approx (2/\gamma)^{1/2} \approx 0.63\times10^{-4}$ rad is the zero-shift angle, $\lambda$ is the wavelength observed on Earth, and $\lambda'$ is the radiation wavelength of the galaxy. Thus a 0.5-μm-wavelength visible light (2.5-eV photon energy) from the galaxy is detected on Earth as wideband radiations, ranging from 1.25 μm (1-eV near infrared) to 0.5 fm (2.5-GeV hard gamma ray).

It is seen from the above example that the red-shifted radiations will have been observed when a distant galaxy approaches us in an extremely high speed, and because of $(\lambda/\lambda')_{\max} \approx \gamma\phi_b^2/2$, the red shift increases as the increase of the galaxy's approaching speed. However the conventional understanding of the red shift has excluded this significant basic result of the principle of relativity.

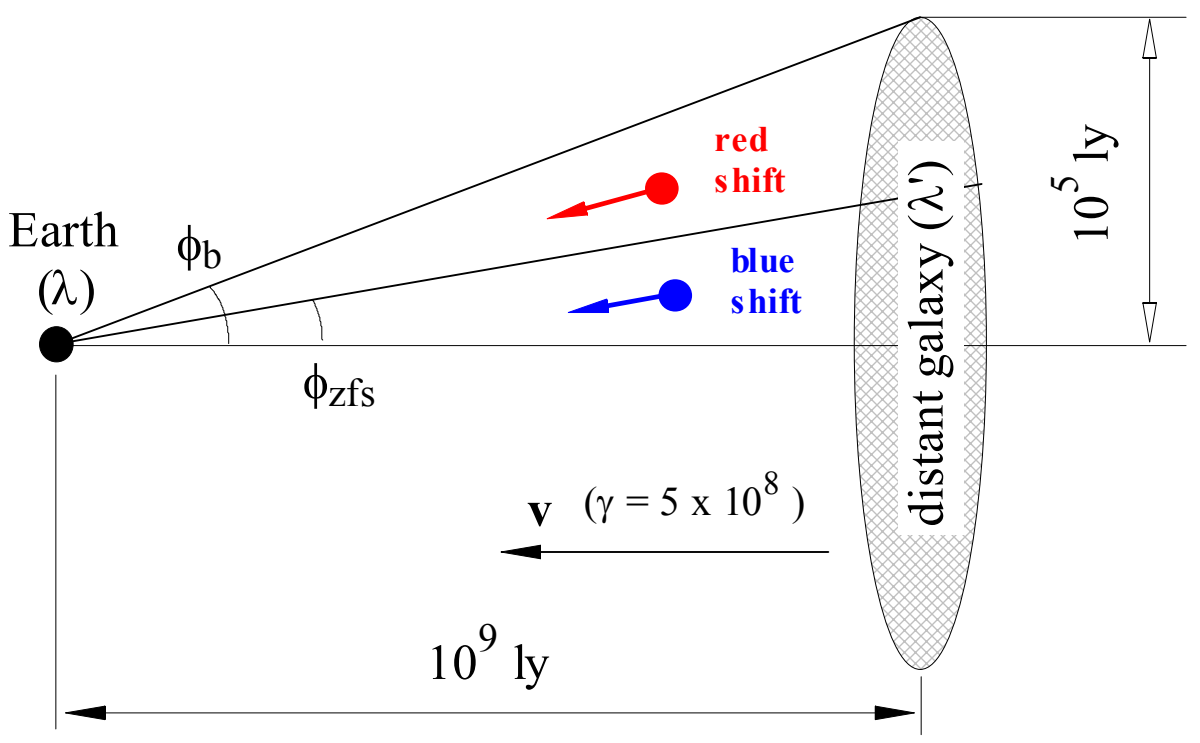

Fig. 7. Illustration for the coexistence of red shift and blue shift from a distant galaxy approaching Earth at a nearly light speed ( $\gamma = 5\times10^8$ ). The oblate revolution-ellipsoid galaxy has a radius of $10^5$ light years and $10^9$ light years away from the Earth. All the electromagnetic radiations with red and blue shifts are distributed within a small angle of $2\phi_b \approx 2\times10^{-4}$ rad, and a 0.5-µm-wavelength visible light from the galaxy is detected on Earth as wideband radiations from 1.25 µm (near infrared) to 0.5 fm (hard gamma ray). Necessary condition for red-shift-for-approaching observation: $\phi_b > \phi_{zfs}$ , namely the galaxy's boundary angle is larger than the zero-shift angle.

In conclusion, we have shown that the relativistic zero-frequency shift in Doppler effect is a strict theoretic result of the special theory of relativity for a plane wave in free space, observed in two inertial frames of relative motion, and in such a case, the definitions of "approaching" and "receding" do not apply. When the zero-shift effect is applied to approximate analysis of a moving point light source, as done by Hovsepyan [11], two unconventional physical implications result. (1) A light source, when it is moving closer to the observer, may cause a red shift. (2) A zero-frequency-shift observation does not necessarily mean that the light source is not moving closer; in contrast, the light source may be moving closer to the observer at a high speed. This fundamental result may provide an alternative way to experimentally examine the principle of relativity, and might have a significant application in astrophysics.

**Note**: After the first version of the paper was put in the arXiv system, Dr. Tomislav Ivezić, to whom the grateful thanks are due, brought to the author's attention Hovsepyan's work [11], where the zero shift or red shift for approaching was first analyzed from the geometrical standpoint. In the present paper, the relativistic zero-frequency shift is shown to be a strict basic result of the special relativity for a plane wave in free space, observed in two inertial frames of relative motion, and a conceptual experiment for verifying the zero shift or red shift for approaching is suggested.

Great appreciation is also extended to the referee for his or her helpful comments and kind suggestions.

## Addendum. Reply to Comment by Sfarti

In the paper [C. Wang, Ann. Phys. (Berlin) **523**, 239 (2011)], I claim that, "A zero-frequency-shift observation does *not necessarily* mean that the light source is not moving closer; in contrast, the light source *may be* moving closer to the observer at a high speed." Unfortunately, this statement is incorrectly interpreted, by Sfarti in his Comment [A. Sfarti, Ann. Phys. **524**, (71) 2012], into what he wanted to mean: the zero-frequency-shift effect is *restricted to* (the light source) "*moving at high speed*". Obviously, Sfarti's Comment made criticism out of thin air. Nevertheless, I would like to thank him for his interest of my work.

COMMENT

# Reply to comment by Sfarti

*Changbiao Wang*

Sfarti [1] argues that, the zero-frequency-shift can be detected at very low $\beta$ values and it is not restricted to "moving at high speed [3]", where his reference [3] refers to my article [2]. However, in my work I do not claim a *restriction to* (the light source) "*moving at high speed*", i.e. my original statement is well distinct from Sfarti's interpretation of it. In [2] movements at high speed are simply mentioned as a possible case for a zero-frequency-shift, but it also allows for other movements. It reads "$\phi_{zfs} < \pi/2$ holds for any $\beta \neq 0$" (with $\varphi_{zfs}$ denoting the zero-shift angle) and later in the text: "… a zero-frequency-shift observation does *not necessarily* mean that the light source is not moving closer; in contrast, the light source *may be* moving closer to the observer at a high speed" [2].

*Changbiao Wang*
*ShangGang Group, 70 Huntington Road, Apartment 11, New Haven, CT 06512, USA*
*E-mail: changbiao_wang@yahoo.com*